\documentclass[preprint,showpacs,showkeys,preprintnumbers,amsmath,amssymb]{revtex4}
\usepackage{graphicx,epsfig}% Include figure files
\usepackage{dcolumn}% Align table columns on decimal point
\usepackage{bm}% bold math

\def\ut#1{\mathop{\vtop{\ialign{##\crcr
     $\hfil\displaystyle{#1}\hfil$\crcr\noalign
     {\kern1pt\nointerlineskip}\hbox{$\hfil\sim\hfil$}\crcr
     \noalign{\kern1pt}}}}}

\begin{document}

\preprint{}

\title{Constraints on a charge in the  Reissner--Nordstr\"om metric for the black hole  at the Galactic
Center

}% Force line breaks
%with \\

\author{Alexander F. Zakharov$^{1,2,3,4,5}$}
\email{zakharov@itep.ru}
 \affiliation{$^{1}$North Carolina Central University, Durham, NC 27707,
 USA\\
% \affiliation{
  $^{2}$Institute of
Theoretical and Experimental Physics, Moscow, 117218, Russia\\
   $^{3}$Joint Institute for Nuclear Research, Dubna, 141980,  Russia\\
   $^{4}$Institute for Computer Aided Design of RAS,     123056, Moscow, Russia\\
   $^{5}$National Research Nuclear University (NRNU MEPHI), 115409,
   Moscow, Russia
}

\date{\today}% It is always \today, today,

\begin{abstract}

 Using an algebraic condition of vanishing discriminant for
multiple roots  of fourth degree polynomials we derive an analytical
expression of a shadow size as a function of a charge in the
Reissner -- Nordstr\"om (RN) metric \cite{Reissner_16,Nordstrom_18}.
We consider shadows for negative tidal charges and charges
corresponding to naked singularities $q=\mathcal{Q}^2/M^2
> 1$, where $\mathcal{Q}$ and $M$ are black hole charge and mass, respectively, with the derived expression.
An introduction of a negative
tidal charge $q$ can describe black hole solutions in theories with
extra dimensions, so following the approach we consider an
opportunity to extend RN metric to negative $\mathcal{Q}^2$, while
for the standard RN metric $\mathcal{Q}^2$ is always non-negative.
We found that for $q
> 9/8$ black hole shadows disappear. Significant tidal charges
$q=-6.4$ (suggested by Bin-Nun
\cite{Bin_Nun10,Bin_Nun10a,Bin_Nun10b}) are not consistent with
observations of a minimal spot size at the Galactic Center observed
in mm-band, moreover, these observations demonstrate that  a
Reissner -- Nordstr\"om black hole with a significant charge $q
\approx 1$ provides a better fit of recent observational data for
the black hole at the Galactic Center in comparison with the
Schwarzschild black hole.

\end{abstract}

%Use showkeys class option if keyword
%                              %display desired

\maketitle

\section{Introduction}
 Soon after discovery of general relativity (GR) first solutions
corresponding to spherical symmetric black holes were found
\cite{Schwarzschild_16,Reissner_16,Nordstrom_18}, however, initially
people were rather sceptical about possible astronomical
applications of the solutions corresponding to black holes
\cite{Einstein_39} (see, for instance, also one of the first
textbooks on GR \cite{Bergmann_42}). Even after an introduction of
the black hole concept by Wheeler \cite{Wheeler_68} (he used the
term in his public lecture in 1967 \cite{Frolov_11}) we don't know
not too many examples where we really need GR models with  strong
gravitational fields which arise near black hole horizons to explain
observational data. The cases where we need strong field
approximation are very important since they give an opportunity to
check GR predictions in a strong field limit, therefore, one could
significantly  constrain of alternative theories of gravity.

One of the most important  options to test a gravity in the strong
field approximation is analysis of relativistic line shape as it was
shown in \cite{Fabian_89}  with assumptions that a line emission is
originated at a circular ring area of a flat accretion disk. Later
on, such signatures of the Fe $K\alpha$-line have been found in the
active galaxy MCG-6-30-15 \cite{Tanaka_95}. Analyzing the spectral
line shape the authors concluded the emission region is so close to
the black hole horizon that one has to use Kerr metric approximation
\cite{Kerr_63} to fit observational data \cite{Tanaka_95}. Results
of our simulations of iron $K\alpha$ line formation are given in
\cite{Zakharov_99} (where we used our approach \cite{Zakharov_94}),
see also \cite{Fabian_10} for a more recent review of the subject.

Now there are two basic observational techniques to investigate a
gravitational potential at the Galactic Center, namely, a)
monitoring the orbits of bright stars near the Galactic Center to
reconstruct a gravitational potential \cite{Ghez_00} (see also a
discussion about an opportunity to evaluate black hole dark matter
parameters in \cite{Zakharov_07} and an opportunity to constrain
some class of an alternative theory of gravity \cite{Dusko_PRD_12});
b) in mm-band with VLBI-technique measuring a size and a shape of
shadows around black hole giving an alternative possibility to
evaluate black hole parameters. The formation of retro-lensing
images (also known as mirage, shadows or "faces" in the literature)
due to the strong gravitational field effects nearby black holes has
been investigated by several authors
\citep{Holz02,Geralico03,ZNDI05,ZNDI05b}.
%\cite{}

Theories with extra dimensions admit astrophysical objects
(supermassive black holes in particular) which are rather different
from standard ones. Tests have been proposed when it would be
possible to discover signatures of extra dimensions in supermassive
black holes since the gravitational field may be different from the
standard one in the GR approach. So, gravitational lensing features
are different for alternative gravity theories with extra dimensions
and general relativity.

Recently, Bin-Nun \cite{Bin_Nun10,Bin_Nun10a,Bin_Nun10b} discussed
an opportunity that the black hole at the Galactic Center is
described by the tidal Reissner-- Nordstr\"om metric which may be
admitted by the Randall--Sundrum II braneworld scenario
\cite{Dadhich_00}. Bin-Nun suggested an  opportunity of evaluating
the black hole metric analyzing (retro-)lensing of bright stars
around the black hole in the Galactic Center. Doeleman et al.
evaluated a size of the smallest spot for the black hole at the
Galactic Center
 with VLBI technique in mm-band \cite{Doeleman_08} (see, constraints done from previous
observations \cite{Shen_05}). Theoretical
 studies showed that the size of the smallest spot near a black hole
 practically coincides  with shadow size because the spot is
 the envelope of the shadow \cite{Falcke00,ZNDI05,ZNDI05b}.
As it was shown \cite{ZNDI05,ZNDI05b}, measurements of the shadow
size around the black hole may help to evaluate parameters of black
hole metric \footnote{One of the first calculations of shapes of
orbits visible by  a distant observer have been done in
\cite{Cunningham_73}. An apparent shape of Kerr black hole was
discussed in \cite{Bardeen_73} (see also a very similar picture in
monograph \cite{chandra}). Later on, the apparent shapes of black
holes are called shadows \cite{Falcke00}.}. Sizes and shapes of
shadows are calculated for different types of black holes and
gravitational lensing in strong gravitational field has been
analyzed in a number of papers \cite{Shadows}.

 We derive an analytic expression for the
black hole shadow size as a function of the tidal charge for the
Reissner-- Nordstr\"om metric. We conclude that observational data
concerning shadow size measurements are not consistent with
significant negative charges, in particular, the significant tidal
charge $q=\mathcal{Q}^{2}/M^2=-6.4$ \footnote{A dimensional tidal
charge $Q=q*M^2=\mathcal{Q}^2$, where $Q$ is a tidal charge,
$\mathcal{Q}$ is the the standard charge in the Reissner--
Nordstr\"om metric. One can consider also negative tidal charges, it
means that a class of metrics with tidal charges is an extension of
the class of the standard Reissner-- Nordstr\"om metrics.},
 discussed in
\cite{Bin_Nun10,Bin_Nun10a,Bin_Nun10b}, where the author used
slightly different notations, namely $q'=q/4$, is practically ruled
out with a very high probability (the tidal charge is roughly
speaking is far beyond $9\sigma$ confidence level). We also show a
smaller shadow sizes in respect to estimates obtained with the
Schwarzschild black hole model can be explained with the Reissner --
Nordstr\"om metric with a significant charge. It was found a
critical $q$ value for shadow existence, namely for $q \leq 9/8$,
Reissner -- Nordstr\"om black holes have shadows while for $q
> 9/8$ the shadows do not exist. Interestingly, the same critical value is responsible for
a qualitative different behavior of quasinormal modes for the
scattering \cite{Chirenti_12} and for existence of circular orbits
of neutral test particles \cite{Pugliese_11}.

As J. A. Wheeler coined "Black holes have no hair": it means that a
black hole is characterized by only three parameters, its mass $M$,
angular momentum $J$ and  charge $\mathcal{Q}$ (see, e.g.
\cite{MTW,Wald84}, or \cite{Hertog_06} for a more recent review).
Therefore, in principle, charged black holes can be formed, although
astrophysical conditions that lead to their formation may look
rather problematic. Nevertheless, one  could not claim that their
existence is forbidden by theoretical or observational arguments.
Moreover, we will show below that observations give a hint about an
existence of a significant charge, but its origin is not clear at
the moment.

Charged black holes are also  object of intensive studies in quantum
gravity, since a static, spherically
 symmetrical  solution of Yang-Mills-Einstein equations with fairly natural
 requirements on asymptotic behavior of the solutions gives a  Reissner-Nordstr\"{o}m
 metric
 \cite{Gal1}. Thus, the metric describes a spherically symmetric
 black hole with a color charge (and (or) a magnetic monopole).
 Later on, color charges have been found for rotating black holes as
 well
 \cite{Kleihaus_97}.

\section{Basic Equations}

The expression for the Reissner - Nordstr\"{o}m metric in natural
units ($G=c=1$) has the form
\begin {equation}
  ds^{2}=-\left(1-\frac{2M}{r}+\frac{\mathcal{Q}^{2}}{r^{2}}\right)dt^{2}+\left(1-\frac{2M}{r}+\frac{\mathcal{Q}^{2}}{r^{2}}\right)^{-1}dr^{2}+
r^{2}(d{\theta}^{2}+{\sin}^{2}\theta d{\phi}^{2}).
\label{RN_Lecce_0}%$
\end {equation}
%where $M$ is the mass of the black hole and $Q$ is its charge.
Applying the Hamilton-Jacobi method to the problem of %describing
geodesics in the Reissner - Nordstr\"{o}m metric, %we note that
the motion of a test particle in the $r$-coordinate can be described
by following equation (see, for example, \cite{MTW})
\begin {eqnarray}
    r^{4}(dr/d\lambda)^{2}=R(r),\label{RN_D 1}
\end {eqnarray}
where $\lambda$ is the affine parameter \cite{MTW} and
\begin {eqnarray}
&&  R(r)=P^{2}(r)-\Delta({\mu}^{2}r^{2}+L^{2}), \nonumber\\
&&  P(r)=Er^{2}-e\mathcal{Q}r,    \label{RN_D_2}\\
&&  \Delta=r^{2}-2Mr+\mathcal{Q}^{2}. \nonumber
\end {eqnarray}

Here, the constants $\mu, E, L$ and $e$ are associated with the
particle, i.e. $\mu$ is its mass, $E$ is energy at infinity, $L$ is
its angular momentum at infinity and $e$ is the particle's charge.

    We shall consider the motion of uncharged particles $(e=0)$ below.
In this case, the expression for the polynomial $R(r)$ takes the
form
\begin {eqnarray}
R(r)=(E^{2}-{\mu}^{2})r^{4}+
2M{\mu}r^{3}-(\mathcal{Q}^{2}{\mu}^{2}+L^{2})r^{2}
+2ML^{2}r-\mathcal{Q}^{2}L^{2}. \label{RN_D_3}
\end {eqnarray}

     Depending on the multiplicities of the roots of the
polynomial $R(r)$,  we can have three types of motion in the $r$ -
coordinate \citep{Zakharov_86}. In particular, by defining the outer
event horizon as usual $r_{+}=M+\sqrt{M^2-\mathcal{Q}^{2}}$
\cite{MTW}, we have:

 (1) if the polynomial $R(r)$
has no roots  for $r\geq r_{+}$,  a test particle is captured by the
black hole;

(2) if $R(r)$ has roots  and $(\partial{R}/\partial{r})(r_{max})\neq
0$  with $r_{max} > r_+$ ($r_{max}$ is the maximal root), a particle
is scattered after approaching the black hole;

(3) if  $R(r)$ has a root  and $R(r_{max}) = (\partial{R}/
\partial{r})(r_{max})=0$,
the particle now takes an infinite proper time to approach the
surface $r = const$.

    If we are considering a photon ($\mu = 0$), its motion in the $r$-coordinate depends
on the root multiplicity of the polynomial $\hat{R}(\hat{r})$
\begin {eqnarray}
\hat{R}(\hat{r})={R(r)}/({M^{4}E^{2}})={\hat{r}}^{4}-\xi^{2}{\hat{r}}^{2}+2{\xi}^{2}\hat{r}
 -{\hat{\mathcal{Q}}}^{2}{\xi}^{2}.\label{RN_D_4}
\end {eqnarray}
where $\hat {r}=r/M, \xi=L/(ME)$ and $\hat
{\mathcal{Q}}=\mathcal{Q}/M.$ Below we do not write the  hat symbol
for these quantities.

One could see from Eq. (\ref{RN_D_4}) and Eqs. (\ref{RN_D_2}) as
well that the black hole charge may influence substantially the
photon motion at small radii ($r \approx 1$), while the charge
effect  is almost negligible at large radial coordinates of photon
trajectories ($r
>> 1$). In the last case we should keep in mind that the charge may cause
only small corrections on photon motion.

\section{Derivation of shadow size as a function of charge}

    Let us consider the problem of the capture cross section of
 a photon by a charged black hole. It is clear that the critical value of the
 impact parameter for a photon to be captured by a Reissner - Nordstr\"{o}m black hole
depends on the multiplicity root condition of the polynomial $R(r)$.
This requirement is equivalent to the  vanishing discriminant
condition \cite{Kostrikin_82}. To find the critical value of impact
parameter for Schwarzschild and RN metrics the condition has been
used for corresponding cubic and quartic equations
\cite{Zakharov_88,Zakharov_91,Zakharov_94b}. In particular, it was
shown that for these cases the vanishing discriminant condition
approach is more powerful in comparison with the procedure excluding
$r_{max}$ from the following system
\begin {eqnarray}
R(r_{max}) = 0, \quad  \\
%\&
\label{RN_D_4a}
%\quad
 \dfrac{\partial{R}}{\partial{r}}(r_{max})=0,
\label{RN_D_4b}
\end {eqnarray}
as it was done, for example,  by Chandrasekhar \cite{chandra} (and
earlier by Darwin \cite{Darwin_58}) to solve similar problems,
because $r_{max}$ is automatically excluded in the condition for
vanishing discriminant.

Introducing the notation $\xi^{2}=l, \mathcal{Q}^{2}=q$, we obtain
\begin {eqnarray}
    R(r)=r^{4}-lr^{2}+2lr-ql.\label{RN_D_5}
\end {eqnarray}
We remind basic algebraic definitions and relations. If we consider
an arbitrary polynomial $f(X)$ with degree $n$
\begin {eqnarray}
f(X) =  X^n+a_1X^{n-1}+...+a_{n-1}X+a_n,
\end {eqnarray}
the elementary symmetric polynomials $s_{k}$ have the following
form, where $X_1,...X_n$ are roots of the polynomial $f(X)$
\cite{Kostrikin_82}
\begin {eqnarray}
s_{k} (X_1,...X_n) =   \sum_{1 \leqslant i_1 < i_2 <...<i_k
\leqslant n} X_{i_{1}}X_{i_{2}}...X_{i_k},
\end {eqnarray}
where $k=1,2,..., n$. The symmetrical $k$-power sum polynomial $p_k$
have the following expression
\begin {eqnarray}
p_{k} (X_1,...X_n) =   X_{1}^k+X_{2}^k+...+X_{n}^k, \quad {\rm for}
\quad k \geq 0.
\end {eqnarray}
To express $p_k$ through $s_k$ one can use Newton's equations
\begin {eqnarray}
p_{k}-p_{k-1}s_1+p_{k-2}s_2+...+(-1)^{k-1}p_1s_{k-1}+(-1)^k ks_k=0,
\quad {\rm for} \quad 1 \leqslant k \leqslant n;
\end {eqnarray}
\begin {eqnarray}
p_{k}-p_{k-1}s_1+p_{k-2}s_2+...+(-1)^{n-1}p_{k-n+1}s_{n-1}+(-1)^n
p_{k-n} s_n=0, \quad {\rm for} \quad  k > n.
\end {eqnarray}
We introduce the following polynomial
\begin {eqnarray}
\Delta_n (X_1,...X_n) =   \prod_{1 \leqslant i < j \leqslant n}
(X_i-X_j),
\end {eqnarray}
which can be represented as the Vandermonde determinant
\begin{eqnarray}
  \Delta_n (X_1,...X_n)=
   \left|
\begin{array}{cccc}
1   &  1 &  ... & 1\\
X_1  &  X_2 & ... & X_n\\
... &  ... &  ... & ...\\
X_1^{n-1}  &  X_2^{n-1} & ... & X_n^{n-1}
\end{array}
\right|.
\end {eqnarray}
According to the discriminant $Dis$ definition we have the
$Dis(s_1,...,s_n)$ polynomial
\begin {eqnarray}
Dis(s_1,...,s_n)=\Delta_n^2 (X_1,...X_n) =   \prod_{1 \leqslant i <
j \leqslant n} (X_i-X_j)^2,
\end {eqnarray}
one can find \cite{Kostrikin_82}
\begin{eqnarray}
  Dis(s_1,...s_n)=
   \left|
\begin{array}{ccccc}
n   &  p_1 & p_2 &  ... & p_{n-1}\\
p_1  &  p_2 & p_3 &... & p_n\\
p_2 & p_3 & p_4  & ... & p_{n+1}\\
... &  ... &  ... & ...& ...\\
p_{n-1}& p_n  &  p_{n+1} & ... & p_{2n-2}
\end{array}
\right|.
\end {eqnarray}

\begin{figure}[th!]
\begin{center}
\includegraphics[width=10.5cm]{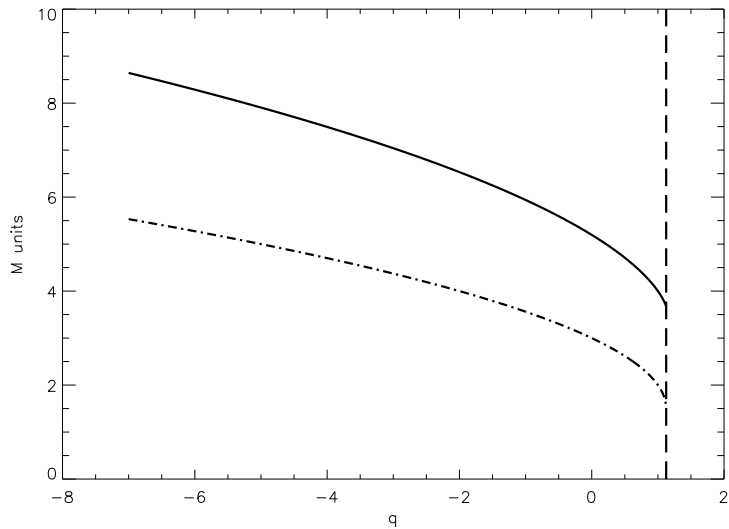}
\end{center}
\caption{Shadow (mirage) radius (solid line) and radius of the last
circular unstable photon orbit (dot-dashed line) in $M$ units as a
function of
 $q$. The critical value $q=9/8$ is shown with dashed vertical line.}
 \label{Fig1}
\end{figure}

Clearly, that the vanishing discriminant condition is equivalent to
an existence of multiple roots among roots $X_1,...X_n$. We apply
this technique for the quartic polynomial $R(r)$ in Eq.
(\ref{RN_D_5}). So that the symmetric $k$-power polynomials for $n =
4$ have the form
\begin {eqnarray}
    p_{k}=X_{1}^{k}+X_{2}^{k}+X_{3}^{k}+X_{4}^{k}, k \geq 0.
\end {eqnarray}
The symmetric elementary polynomials for $n = 4$ have the form
\begin {eqnarray}
&&  s_{1}=X_{1}+X_{2}+X_{3}+X_{4},                                         \nonumber\\
&&  s_{2}=X_{1}X_{2}+X_{1}X_{3}+X_{1}X_{4}+X_{2}X_{3}+X_{2}X_{4}+X_{3}X_{4},\nonumber\\
&&  s_{3}=X_{1}X_{2}X_{3}+X_{2}X_{3}X_{4}+X_{1}X_{3}X_{4}+X_{1}X_{2}X_{4},                   \\
&&  s_{4}=X_{1}X_{2}X_{3}X_{4}. \nonumber
\end {eqnarray}
We calculate the discriminant of the family $X_{1},X_{2},X_{3},X_{4}$\\
%\begin{displaymath}
\begin {eqnarray}
%\begin {array}{cccc}
   Dis(s_{1},s_{2},s_{3},s_{4})=
   \left|
\begin{array}{cccc}
%\left|
1& 1& 1& 1\\
X_{1}& X_{2}& X_{3}& X_{4}\\
X_{1}^{2}& X_{2}^{2}& X_{3}^{2}& X_{4}^{2}\\
X_{1}^{3}& X_{2}^{3}& X_{3}^{3}& X_{4}^{3}
%\right|
\end {array}
\right|^2 =\left|
\begin{array}{cccc}
%\left|
4   &  p_{1}& p_{2}& p_{3}\\
p_{1}& p_{2}& p_{3}& p_{4}\\
p_{2}& p_{3}& p_{4}& p_{5}\\
p_{3}& p_{4}& p_{5}& p_{6}
\end {array}
\right|
%$
\end {eqnarray}
%\end{displaymath}
Expressing the polynomials $p_{k} (1\leq k \leq 6)$ in terms of the
polynomials $s_{k} (1\leq k\leq 4)$ and using Newton's equations we
calculate the polynomials and discriminant of the family
$X_{1},X_{2},X_{3},X_{4}$ in roots of
 the polynomial $R(r)$; we obtain
\begin {eqnarray}
&&  p_{1}=s_{1}=0, \quad p_{2}=-2s_{2}, \quad p_{3}=3s_{3},\nonumber\\
&& p_{4}=2s_{2}^{2}-4s_{4}, \quad p_{5}=-5s_{3}s_{2},\\
&& p_{6}=-2s_{2}^{3}+3s_{3}^{2}+6s_{4}s_{2},\nonumber
\end {eqnarray}
where $s_{1}=0,s_{2}=-l,s_{3}=-2l,s_{4}=-ql,$ corresponding to the
polynomial $R(r)$ in Eq. (\ref{RN_D_5}). The discriminant $Dis$ of
the polynomial $R(r)$ has the form:
\begin {eqnarray}
   Dis(s_{1},s_{2},s_{3},s_{4})=
   \left|
\begin{array}{cccc}
4   &  0 &   2l & -6l\\
0  &  2l &  -6l & 2l(l+2q)\\
2l &  -6l &  2l(l+2q) & -10l^{2}\\
-6l &  2l(l+2q) & -10l^{2} & 2l^{2}(l+6+3q)
\end{array}
\right|=\nonumber\\
=16l^{3}[l^{2}(1-q)+l(-8q^{2}+36q-27)-16q^{3}]. \label{RN_D_6}
\end {eqnarray}
The polynomial $R(r)$ thus has a multiple root  if and only if
\begin {eqnarray}
  l^{3}[l^{2}(1-q)+l(-8q^{2}+36q-27)-16q^{3}]=0. \label{RN_D_7}
\end {eqnarray}
Excluding the case $l=0$, which corresponds to a multiple root at
$r=0$, we find
 that the polynomial $R(r)$ has a multiple root for $ r\geq r_{+}$ if and only if
\begin {eqnarray}
 l^{2}(1-q)+l(-8q^{2}+36q-27)-16q^{3}=0. \label{RN_D_8}
\end {eqnarray}
If $q=0$, we obtain the well-known result for a Schwarzschild black
hole \citep{MTW,Wald84,Lightman}, $l_{\rm cr}=27$, or
$\xi_{cr}=3\sqrt{3}$ (where $l_{\rm cr}$ is the positive root of Eq.
(\ref{RN_D_8})). If $q=1$, then $l = 16$, or $\xi_{cr}=4$, which
also corresponds to numerical results given in paper \cite{Young}.
    The photon capture cross section for an  extreme charged
    black hole  turns out to be considerably smaller than the capture cross section of a
 Schwarzschild black hole. The critical value of the impact parameter,
 characterizing the capture cross section for a RN black hole, is determined by the equation
\begin {eqnarray}
l_{\rm cr}=\frac{(8q^{2}-36q+27)+\sqrt{D_1}}{2(1-q)}, \label{RN_D_9}
\end {eqnarray}
where
$D_1=(8q^{2}-36q+27)^{2}+64q^{3}(1-q)=-512\left(q-\dfrac{9}{8}\right)^3.$
It is clear from the last relation that there are circular unstable
photon orbits only for $q \le \dfrac{9}{8}$ (see also results in
\cite{Pugliese_11} about the same critical value). Substituting
Eq.(\ref{RN_D_9}) into the expression for the coefficients of the
polynomial $R(r)$ it is easy to calculate the radius of the unstable
circular photon orbit (which is the same as
 the minimum periastron distance). The orbit of a
 photon moving from infinity with the critical impact parameter, determined
in accordance with Eq.(\ref{RN_D_9}) spirals into  circular orbit.
To find a radius of photon unstable orbit we will solve Eq.
(\ref{RN_D_4b}) substituting $l_{\rm cr}$ in the relation. From
trigonometric formula for roots of cubic equation we have
\begin {eqnarray}
r_{\rm crit}=2\sqrt{\frac{l_{\rm cr}}{6}} \cos{\frac{\alpha}{3}},
\label{RN_D_10}
\end {eqnarray}
where
\begin {eqnarray}
\cos \alpha={-\sqrt{\frac{27}{2 l_{\rm cr}}}}, \label{RN_D_11}
\end {eqnarray}
As it was explained in \cite{ZNDI05b} this leads to the formation of
shadows  described by the critical value of $\xi_{cr}$ or, in other
words, in the spherically symmetric case, shadows are circles with
radii $\xi_{cr}$. Therefore, by measuring the shadow size, one could
evaluate the black hole charge in  black hole mass units $M$. In
Fig. \ref{Fig1} a shadow radius and a radius of last unstable orbit
for photons as a function of $q$ are given as a function of charge
(including possible tidal charge with a negative $q$ and
super-extreme charge $q>1$).

%\begin{figure}[th!]
%\begin{center}
%\includegraphics[width=10.5cm]{RN_4_unstable_orbits_revision.ps}
%\end{center}
%\caption{Radius of the last circular unstable photons orbit in $M$
%units as a function of
% $q$.}
% \label{Fig2}
%\end{figure}
%\newpage
%\\
%\begin{eqnarray}

%\end{eqnarray}

\section{Consequences}

\subsection{A disappearance of shadows for naked singularities}

In spite of the cosmic censorship hypothesis \cite{Penrose_69} that
a singularity has to be shielded by a horizon, properties of naked
singularities  are a subject of intensive theoretical studies. As
usual spherical symmetrical cases are easier for  analysis and RN
metrics with super extreme charge $q>1$ are investigated in a number
of papers, see, for instance \cite{Virbhadra_02} and references
therein.

So, if we assume that $q> 1$, we can see from Eq. (\ref{RN_D_9})
that for $q \leq 9/8$ we have shadows, while for $q
> 9/8$ the shadows do not exist. For these charges ($q
> 9/8$)
incoming photons always scattering by black holes for $l \neq 0$
because the polynomial $R(r)$ has no multiple roots but it has a
single positive root (it  means scattering) since for great positive
$r$ we have $R(r)
> 0$ while $R(0) <0$. The degenerate case of radial trajectories of
photons ($l = 0$) can be ignored as the case with "zero measure" or
the structural unstable case using the Poincar\'e -- Pontryagin --
Andronov -- Anosov -- Arnold terminology \cite{Arnold_88}. It means
that in any small vicinity a behavior of other geodesics from the
radial ones is qualitatively different, therefore, such objects can
not be observed in nature. Therefore, shadows exist only for $q \leq
9/8$. So,  $q=9/8$ is critical value which is characterized
"catastrophe" \cite{Arnold_04} or the qualitatively different
behavior of the system (the appearance and the disappearance of
shadows).

For the critical $q=9/8$ we have the smallest shadow  with $l=27/2$
and a shadow size $\xi=\sqrt{13.5} \approx 3.674$ (in $M$ units) or
$37.5~\mu as$ in diameter for the black hole at the Galactic Center.
For this impact parameter we have corresponding circular unstable
orbit for photons with $r=1.5$ (in $M$ units).

\subsection{Observational constraints on a charge of the black hole
at the Galactic Center}

{ If we adopt the distance toward the Galactic Center $d_*=8.3\pm
0.4$~kpc (or $d_*=8.35\pm 0.15$~kpc \cite{Reid_14}) and mass of the
black hole $M_{BH}=(4.3\pm 0.4) \times 10^6 M_\odot$
\cite{Gillessen_09,Falcke_13} (a significant part of black hole mass
uncertainty is connected with a distance determination uncertainty
\cite{Falcke_13}), then we have the angle 10.45~$\mu as$ for the
corresponding Schwarzschild radius $R_g=2.95*
\dfrac{M_{BH}}{M_\odot}*10^5$~cm roughly with 10\% uncertainty of
black hole mass and distance estimations, so a shadow size for the
Schwarzschild black hole is around 53~$\mu as$, for a black hole
with a tidal charge ($q=-6.4$) suggested by Bin-Nun
\cite{Bin_Nun10,Bin_Nun10a,Bin_Nun10b} a shadow size is about
86.1~$\mu as$, while for the extreme charge ($q=1$) and critical
charge ($q=9/8$) the shadow sizes are 40.9~$\mu as$ and 37.5~$\mu
as$, respectively.Uncertainties of angular shadow size evaluations
are at a level around 10\% which corresponds to an
 uncertainty of black hole mass evaluation.

}

\subsection{Comparison with observations}

A couple of year ago Doeleman et al. \cite{Doeleman_08} claimed that
intrinsic diameter of Sgr $A^*$ is $37^{+16}_{-10}~\mu as$ at the
$3~ \sigma$ confidence level. If we believe in GR and the central
object is a black hole, then we have to conclude that a shadow
practically coincides with the intrinsic diameter, so in spite of
the fact that a Schwarzschild black hole is marginally consistent
with observations, a Reissner -- Nordstr\"om black hole provides
much better fit of a shadow size, while a black hole with a
significant tidal charge ($q=-6.4$) is out of a more 9~$\sigma$
level interval. Later on,  an accuracy of intrinsic size
measurements was significantly improved,  so  Fish et al.
\cite{Fish_11} gave $41.3^{+5.4}_{-4.3}~\mu as$ (at 3~$\sigma$
level) on day 95, $44.4^{+3.0}_{-3.0}~\mu as$ on day 96 and
$42.6^{+3.1}_{-2.9}~\mu as$ on day 97, so a tidal charge ($q=-6.4$)
is out of $26~\sigma$ level for day 95 and even less probable for
other observations.

The black hole in the elliptical galaxy M87 looks also perspective
to evaluate shadow size \cite{Doeleman_12} (probably  even its shape
in the future to estimate a black hole spin) because the distance
toward the galaxy is $16 \pm 0.6 $~Mpc \cite{Blakeslee_09}, black
hole mass is $M_{M87}=(6.2\pm 0.4) \times 10^9 M_\odot$
\cite{Gerbhardt_11}, so that an angle $(7.3 \pm 0.5) \mu as$
corresponds to the Schwarzschild radius \cite{Doeleman_12}, so the
angle is comparable with the corresponding value considered earlier
for our Galactic Center case. Recently, it was reported that
smallest shadow size is $5.5 \pm 0.4  R_{SCH}$ with $1~\sigma$
errors  (where $R_{SCH}=2 G M_{M87}/c^2$) \cite{Doeleman_12}, so
that at the moment the shadow size is consistent with the
Schwarzschild metric for the object.

\section{Conclusions}

Based on observations \cite{Doeleman_08,Fish_11} one can say that
for the Schwarzschild black hole model we have tensions between
evaluations of black hole mass done with observations of bright star
orbits near the Galactic Center and the evaluated shadow size. To
reduce tensions between estimates of the black hole mass and the
intrinsic size measurements, one can use the Reissner -- Nordstr\"om
metric with a significant charge which is comparable with the
critical one. We do not claim that the corresponding charge has an
electric origin because an interstellar environment is electrically
neutral, so the corresponding charge may be induced (like a tidal
charge induced by extra dimension) and has a non-electric origin.
Charge estimates for the Reissner -- Nordstr\"om metric given from
geodesic trajectories for orbital motions are given in
\cite{Iorio_12}.

Recent estimates of the smallest structure in the M87 published in
paper \cite{Doeleman_12} do not need an introduction of charge
(tidal or normal) to fit observational data because sizes of the
smallest spot near the black hole at the object are still consistent
with the shadow size evaluated for the Schwarzschild metric.

\begin{acknowledgments}

The author thanks D. Borka, V. Borka Jovanovi\'c, F. De Paolis, G.
Ingrosso, P. Jovanovi\'c, S. M. Kopeikin,  A. A. Nucita, B. Vlahovic
for useful discussions. The authors acknowledges also A. Broderick
and C. L\"ammerzahl for conversations at GR-20/Amaldi-10 in Warsaw.
The author thanks anonymous referees for  their very useful and
constructive remarks and J. Nimmo (NCCU) for his kind help to
improve English of the manuscript.

 The work was supported in part by the NSF
(HRD-0833184) and NASA (NNX09AV07A) at NASA CADRE and NSF CREST
(NCCU, Durham, NC, USA)  and RFBR 14-02-00754a at ICAD of RAS
(Moscow).
\end{acknowledgments}

\end{document}